\documentclass[aip,jap,floatfix,showpacs,reprint]{revtex4-1}
\usepackage{amsmath}
\usepackage{graphicx}

\begin{document}

\title{Geometrical effects on spin injection: 3D spin drift diffusion model}
\author{Juzar Thingna}
\thanks{Author to whom correspondence should be addressed. Electronic mail: juzar@nus.edu.sg}
\author{Jian-Sheng~Wang}
\affiliation{Department of Physics and Centre for Computational Science and Engineering, National University of Singapore, Singapore 117542, Republic of Singapore}
\date{\today}

\begin{abstract}
We discuss a three-dimensional (3D) spin drift diffusion (SDD) model to inject spin from a ferromagnet (FM) to a normal metal (N) or semiconductor (SC). Using this model we investigate the problem of spin injection into isotropic materials like GaAs and study the effect of FM contact area and SC thickness on spin injection. We find that in order to achieve detectable spin injection a small contact area or thick SC samples are essential for direct contact spin injection devices. We investigate the use of thin metal films (Cu) proposed by S.B. Kumar et~al. and show that they are an excellent substitute for tunnelling barriers (TB) in the regime of small contact area.  Since most tunnelling barriers are prone to pinhole defects, we study the effect of pinholes in AlO tunnelling barriers and show that the reduction in the spin-injection ratio ($\gamma$) is solely due to the effective area of the pinholes and there is no correlation between the number of pinholes and the spin injection ratio.
\end{abstract}

\maketitle

\section{Introduction}
Spintronics has been a rapidly growing field from the past few years. The main interest in spintronics has arisen because of a spin-interference device proposed by Datta and Das \cite{datta and das}, which is based on spin precession controlled by a gate voltage. Since then a lot of modified devices have been proposed, but in all these devices the main problem is of efficient spin injection from the ferromagnet (FM) to the normal metal (N) or semiconductor (SC). In order to understand efficient spin injection it is necessary to understand the coupling between charge and spin currents, which was first described by Aronov \cite{aronov} and later developed by Johnson and Silsbee in terms of thermodynamic processes \cite{johnson and silsbee}. P.C. van Son et~al. \cite{van son} later proposed a much simpler linear response model based on spin drift and diffusion (SDD) to describe transport across FM-N/SC interfaces. SDD model was successfully applied to current perpendicular to plane geometries of giant magneto-resistance by Valet and Fert and they also established the connection between the diffusive model and the Boltzmann equation \cite{valet and fert}. The problem in injecting spin from FM to SC is due to the difference in the conductivities and spin diffusion lengths of the materials and is termed as the conductivity mismatch problem \cite{schmidt}, which was solved by Rashba using spin dependent boundary resistance \cite{rashba1}.

Although SDD has been used to describe spin transport across many local \cite{rashba2, bala} and non-local geometries \cite{jedema1, kimura1, jedema2, jedema3}, it is applied as a 1D theory. It was first shown by T. Kimura et~al. that a 1D theory is insufficient to describe a 3D experimental geometry \cite{kimura2}. A 2D extension of the SDD model by Ichimura et~al. has been used to study the spatial variations of the spin current and the electrochemical potential\cite{ichimura} and also a quasi 3D model based on spin dependent resistive elements (SDRE) was proposed by Hamrle et~al., where each SDRE follows 1D equations\cite{hamrle}. Although these models are extensions to the 1D case they are still insufficient to take into account all the effects encountered in an experimental 3D device. In this paper we will first describe the SDD model in 3D with the appropriate boundary conditions and point out the shortcomings of the 1D model. We then apply our model to describe the effects of SC height and the FM contact area on the spin-injection efficiency of the device. Here we will discuss key issues such as the possibility of direct contact spin injection device and the use of thin metal films as substitutes for tunnelling barriers. Our main findings are that when device parameters are smaller than the order of the spin diffusion length, the spin injection ratio is influenced dramatically which is not predicted by the simplistic 1D model. Finally, we will discuss pinhole defects in tunnelling barriers and the effect of pinholes on the spin injection ratio of the device.

\section{Theory}
In order to derive the spin drift diffusion equations we assume that far from the interface at temperatures lower than the Curie temperature most scattering events will conserve the spin direction and thus the spin up and spin down electrons will flow almost independently of each other \cite{mott}. Also if the spin scattering occurs at much longer time-scale than other electron scattering events we can define the electrochemical potentials $\mu_{\scriptscriptstyle \uparrow}$ and $\mu_{\scriptscriptstyle \downarrow}$ for both the spin channels. Thus in the linear response regime the current carried by the spin-up ($\mathbf{j}_{\scriptscriptstyle \uparrow}$) and spin-down ($\mathbf{j}_{\scriptscriptstyle \downarrow}$) channel is given by Ohm's law:
\begin{equation}
\mathbf{j}_{\scriptscriptstyle \uparrow,\scriptscriptstyle \downarrow}=\frac{\sigma_{\scriptscriptstyle \uparrow,\scriptscriptstyle \downarrow}}{e} \vec{\nabla} \mu_{\scriptscriptstyle \uparrow,\scriptscriptstyle \downarrow} ,
\label{eqn:1}
\end{equation}
where, $\sigma_{\scriptscriptstyle \uparrow,\scriptscriptstyle \downarrow}=\sigma (1\pm\alpha)/2$ is the spin dependent electrical conductivity\cite{note0} and $e$ ($>$ 0) is the electron charge. 
Near the interface the spin can diffuse from the up-spin channel to the down-spin channel and thus the coupling of the two spin channels is given by the diffusion equation:
\begin{equation}
\frac{\mu_{\scriptscriptstyle \uparrow}-\mu_{\scriptscriptstyle \downarrow}}{\tau}=D\nabla^{2}(\mu_{\scriptscriptstyle \uparrow}-\mu_{\scriptscriptstyle \downarrow}) ,
\label{eqn:2}
\end{equation}\\
where, $D$ is the Diffusion constant and $\sqrt{D\tau}=\lambda$, the spin diffusion length.

Now in order to simplify notation we use the following transformations \cite{rashba2}
\begin{equation}
\zeta=\mu_{\scriptscriptstyle \uparrow}-\mu_{\scriptscriptstyle \downarrow} ,
\label{eqn:3}
\end{equation}
\begin{equation}
Z=\frac{\mu_{\scriptscriptstyle \uparrow}+\mu_{\scriptscriptstyle \downarrow}}{2} ,
\label{eqn:4}
\end{equation}
\begin{equation}
\Gamma=\frac{(\mathbf{j_{\scriptscriptstyle \uparrow}}-\mathbf{j_{\scriptscriptstyle \downarrow}})\cdot\hat{n}_{\scriptscriptstyle 1}}{(\mathbf{j_{\scriptscriptstyle \uparrow}}+\mathbf{j_{\scriptscriptstyle \downarrow}})\cdot\hat{n}_{\scriptscriptstyle 1}} ,
\label{eqn:5}
\end{equation}
where $\Gamma$ is the spin-injection ratio and $\hat{n}_{\scriptscriptstyle 1}$ is taken as the normal to the surface along the flow direction.

Eq.~(\ref{eqn:1}) and (\ref{eqn:2}) then transform into:
\begin{equation}
\nabla^{2}\zeta=\frac{\zeta}{\lambda^{2}} ,
\label{eqn:6}
\end{equation}
\begin{equation}
\vec{\nabla} Z=-\left(\frac{\Delta \sigma}{2\sigma}\right)\vec{\nabla} \zeta + \frac{\mathbf{J}e}{\sigma} ,
\label{eqn:7}
\end{equation}
\begin{equation}
\Gamma = \frac{2\sigma_{\scriptscriptstyle \uparrow} \sigma_{\scriptscriptstyle \downarrow}}{\sigma}\frac{(\vec{\nabla} \zeta)\cdot\hat{n}_{\scriptscriptstyle 1}}{\mathbf{J}\cdot\hat{n}_{\scriptscriptstyle 1} e}+\frac{\Delta \sigma}{\sigma} ,
\label{eqn:8}
\end{equation}
where, $\Delta \sigma=\sigma_{\scriptscriptstyle \uparrow}-\sigma_{\scriptscriptstyle \downarrow}$ and $\mathbf{J}=\mathbf{j}_{\scriptscriptstyle \uparrow}+\mathbf{j}_{\scriptscriptstyle \downarrow}$ is the total current through the system.

The boundary conditions for the transformed equations are\cite{note1}:
\begin{equation}
\zeta_{\scriptscriptstyle N}|_{\scriptscriptstyle 0}-\zeta_{\scriptscriptstyle F}|_{\scriptscriptstyle 0}=2 r_{\scriptscriptstyle c}\left(\Gamma-\frac{\Delta \Sigma}{\Sigma}\right)(\mathbf{J}\cdot\hat{n}_{\scriptscriptstyle 1})e ,
\label{eqn:9}
\end{equation}
\begin{equation}
\zeta|_{\scriptscriptstyle \pm \scriptscriptstyle \infty}=0 ,
\label{eqn:10}
\end{equation}
\begin{equation}
Z_{\scriptscriptstyle N}|_{\scriptscriptstyle 0}-Z_{\scriptscriptstyle F}|_{\scriptscriptstyle 0}=r_{\scriptscriptstyle c}\left(1-\frac{\Delta \Sigma}{\Sigma}\Gamma\right)(\mathbf{J}\cdot\hat{n}_{\scriptscriptstyle 1})e ,
\label{eqn:11}
\end{equation}
\begin{equation}
\Gamma_{\scriptscriptstyle F}|_{\scriptscriptstyle 0}=\Gamma_{\scriptscriptstyle N}|_{\scriptscriptstyle 0} ,
\label{eqn:12}
\end{equation}
\begin{equation}
(\mathbf{j}_{\scriptscriptstyle \uparrow}-\mathbf{j}_{\scriptscriptstyle \downarrow})\cdot\hat{n}_{\scriptscriptstyle 2}=0 , 
\label{eqn:13}
\end{equation}
where,$\Delta\Sigma=\Sigma_{\scriptscriptstyle \uparrow}-\Sigma_{\scriptscriptstyle \downarrow}$, $\Sigma=\Sigma_{\scriptscriptstyle \uparrow}+\Sigma_{\scriptscriptstyle \downarrow}$, $r_{\scriptscriptstyle c}=\Sigma/(4\Sigma_{\scriptscriptstyle \uparrow}\Sigma_{\scriptscriptstyle \downarrow})$ is the effective contact resistance and $\hat{n}_{\scriptscriptstyle 2}$ is the normal to the boundary of the domain. Here subscript $0$ denotes the interface and $\Omega$ is the domain of the device\cite{note2}. In addition to Eq.~(\ref{eqn:9}), (\ref{eqn:10}), (\ref{eqn:11}), (\ref{eqn:12}) which are generally used for the 1D model, we use the Eq.~(\ref{eqn:13}) for the 3D model. The additional boundary condition ensures that no spin current leaks out of the device. 

We solve Eq.~(\ref{eqn:6}) and (\ref{eqn:7}) using the program FF3D\cite{ff3d}, which employs fictitious domain finite element method (FEM). Fictitious domain method allows us to change the device geometry without actually changing the grid, hence in the case of defects in tunnelling barriers this turns out to be the most viable option. In order to obtain the spin injection ratio we solve the equations in the two regions (FM and SC/N) separately and iteratively vary the boundary conditions till a convergence of 10$^{-6}$ in the spin injection ratio is achieved. To ensure that all the results are well converged with respect to the grid parameters we make sure that for boundaries where Eq.~(\ref{eqn:10}) needs to be satisfied the boundary is $\sim$ 5 $\times$ spin diffusion length. Also the grid spacing is varied until convergence is achieved. As an additional check we use the converged grid parameters for 1D geometries to ensure that the theoretical results are recovered.
\begin{figure}
\includegraphics[width=8.5cm, height=6.0cm]{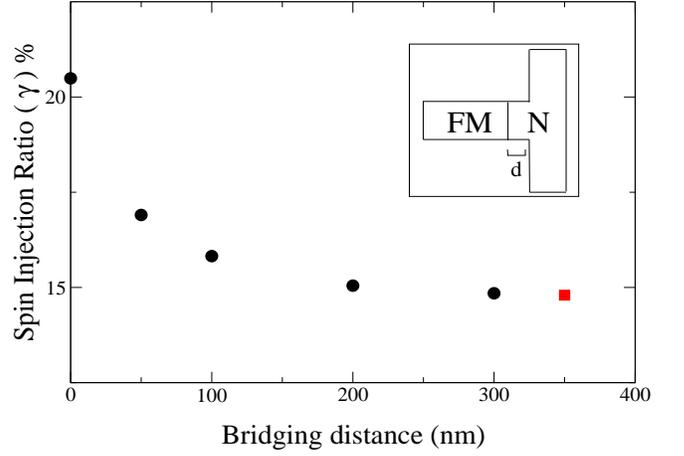}
\caption{(Color Online) Spin injection ratio as a function of the bridging distance $d$ for a three-terminal device. Square (red-online) represents the 1D result.}
\label{fig:3_terminal}
\end{figure}

We first consider a three-terminal, NiFe (FM) and Cu (N), device with the following transport parameters\cite{bala}: $\sigma_{\scriptscriptstyle{\rm{NiFe}}} = 8.62\times 10^{6} /\Omega$m, $\sigma_{\scriptscriptstyle{\rm{Cu}}} = 59.52\times 10^{6} /\Omega$m, $\lambda_{\scriptscriptstyle{\rm{NiFe}}} = 10$ nm, $\lambda_{\scriptscriptstyle{\rm{Cu}}} = 140$ nm, $\alpha_{\scriptscriptstyle{\rm{NiFe}}} = 0.4$, $\alpha_{\scriptscriptstyle{\rm{Cu}}} = 0$ and $r_{\scriptscriptstyle c} = 0$ $\Omega$m$^{2}$. For such a device the spin current is injected using the ferromagnet and detected in the normal metal. For this device we measure the amount of spin current injected into the normal metal as a function of the bridging distance ($d$) between the ferromagnet and the normal metal. According to the 1D SDD model the spin injection ratio ($\gamma$) is given by, 
\begin{equation}
\gamma=\frac{\alpha}{\frac{\lambda_{\scriptscriptstyle N}\sigma_{\scriptscriptstyle F}(1-\alpha^{2})}{\lambda_{\scriptscriptstyle F}\sigma_{\scriptscriptstyle N}}+1} ,\\
\label{eqn:14}
\end{equation}
\begin{figure*}[t]
\includegraphics[width=15cm, height=4.0cm]{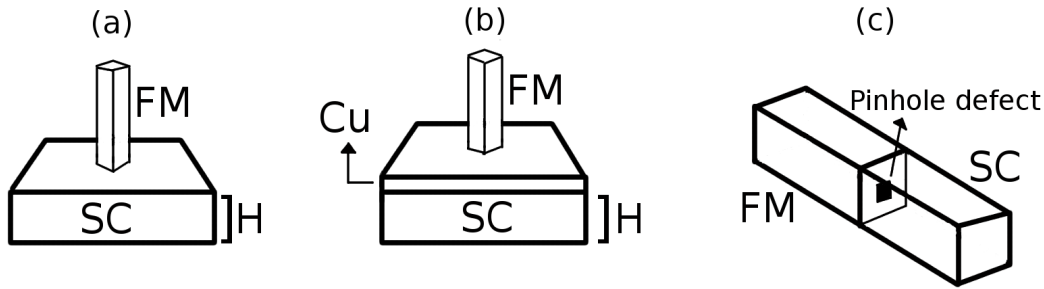}
\caption{(Color Online) Schematic representation of the various device geometries.Fig (a) and (b) represent the device geometry with varying contact area and thickness (H). Fig (c) represents the device geometry used to study the pinhole defects in tunnelling barriers.}
\label{fig:devices}
\end{figure*}
which does not depend on the bridging distance. Plot in Fig.~\ref{fig:3_terminal} shows $\gamma$ as a function of the bridging distance for a more general 3D device (Fig.~\ref{fig:3_terminal} inset). It can be clearly seen that the 1D model is recovered in the limiting case where $d > \lambda_{N}$ (square, red-online). As $d$ decreases $\gamma$ increases rapidly and thus in order to achieve a better spin injection ratio the ferromagnet should be connected close to the normal metal. It should also be noted that in the 1D model the only device dimension is along the flow of current and hence the 1D model is quite limited.

%
\begin{figure}[t]
\includegraphics[width=8.5cm, height=6.0cm]{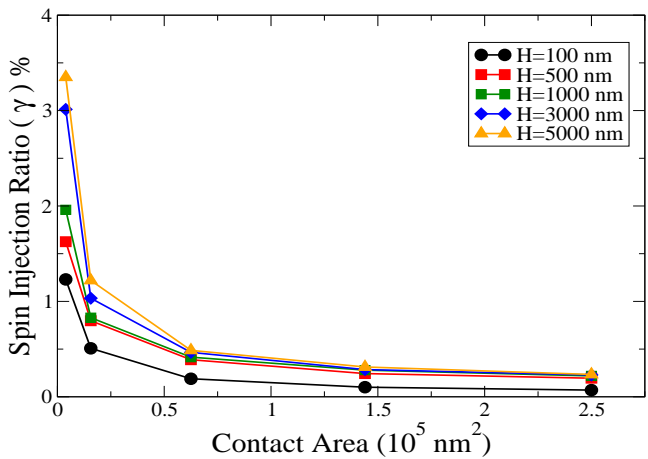}
\caption{(Color Online) Spin injection ratio as a function of the contact area for direct contact device. Different curves are for different semiconductor heights.}
\label{fig:CA}
\end{figure}
\begin{figure}[t]
\includegraphics[width=8.5cm, height=6.0cm]{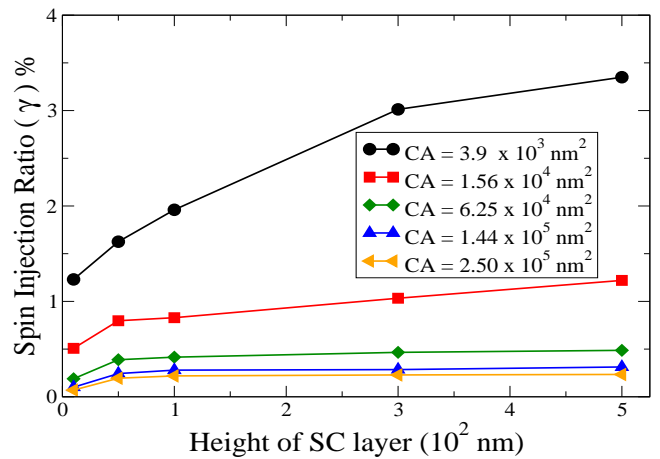}
\caption{(Color Online) Spin injection ratio as a function of semiconductor height for a direct contact device. Different curves are for different contact areas.}
\label{fig:H}
\end{figure}
\section{Results and Discussions}
In order to show the potential of 3D modelling we study the effect of height (H) of the semiconductor and contact area (CA) [see Fig.~\ref{fig:devices} (a)] on n-GaAs with the following transport parameters: $\sigma_{\scriptscriptstyle{\rm{n-GaAs}}} = 10^{5} /\Omega$m, $\lambda_{\scriptscriptstyle{\rm{n-GaAs}}} = 1$ $\mu$m, $\alpha_{\scriptscriptstyle{\rm{n-GaAs}}} = 0$ and $r_{\scriptscriptstyle c} = 0$ $\Omega$m$^{2}$. Kumar et~al. \cite{bala} have shown the effect of nano-pillar ferromagnet on the spin injection ratio by incorporating the effects of spreading resistance in the 1D model, they could not show the effect of SC height since the model was essentially 1D. It should also be noted that they used a contact area of $\sim 12.5$ nm$^{2}$ which is extremely difficult to reproduce experimentally and hence we study the effect of varying contact area on the spin injection ratio (Fig.~\ref{fig:CA}). It can be seen from Fig.~\ref{fig:CA} that as the contact area increases $\gamma$ decreases rapidly. Due to the rapid decay a contact area of at least $\sim 10^{3}$ nm$^{2}$ would be required in order to achieve a direct contact spin signal into the device. According to our knowledge there has been no experimental evidence of direct contact spin injection into a semiconductor because most experimental geometries have a contact area of $\sim 10^{5}$ nm$^{2}$. We also study the effect of sample height on the spin injection ratio (Fig.~\ref{fig:H}). For a given contact area the spin injection ratio increases nearly linearly for small sample thickness, but as the sample thickness approaches the order of $\lambda_{\scriptscriptstyle SC}$ the rate of change of the spin injection ratio reduces drastically. Thus in order to achieve better spin injection in direct contact devices a smaller contact area and a sample height $\gg \lambda_{\scriptscriptstyle SC}$ are required.
\begin{figure}[t]
\includegraphics[width=8.5cm, height=6.0cm]{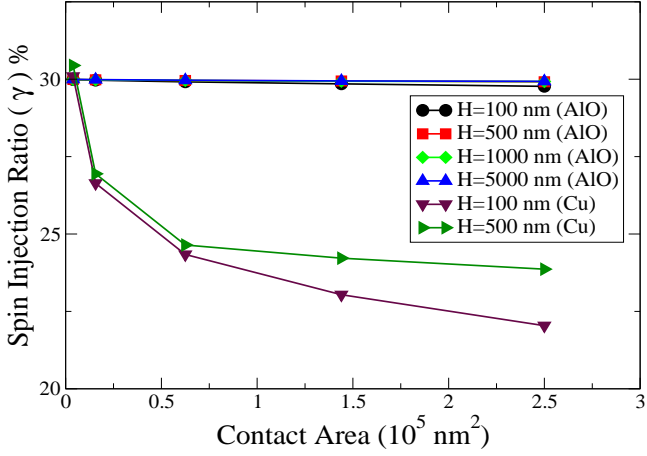}
\caption{(Color Online) Spin injection ratio as a function of the contact area for AlO tunnelling barriers and Cu buffer layer for various semiconductor heights.}
\label{fig:TB}
\end{figure}
Since extremely small contact areas are not feasible experimentally we investigate the role of tunnelling barrier (AlO) and a thin metal (Cu) layer insertion for device geometry shown in Fig.~\ref{fig:devices} (b). The parameters for the AlO tunnelling barrier are:\cite{fert and jaffres} $r_{\scriptscriptstyle c} = 10^{-7}$ $\Omega$m$^{2}$ and $\Delta \Sigma/\Sigma = 0.3$ and the thickness of the Cu insertion layer used is 50 nm. Tunnelling barriers are an excellent solution to the conductivity mismatch problem, but one of the main practical problems with tunnelling barriers is pinhole defects \cite{dlubak, tombros}. If we consider ``ideal'' tunnelling barriers without any pinholes then Fig.~\ref{fig:TB} shows that they are robust to variations in contact area and N/SC height. But for small contact areas we can see that thin Cu films ($\sim$ 50 nm) \cite{note3} can be excellent substitutes for tunnelling barriers. Here $\gamma$ is measured at the Cu-SC interface, taking into account the spin relaxation within the Cu buffer layer. It should be noted here that although using thin Cu film will decrease the effective spin diffusion length of the device, thin metal films don't have defect problems like tunnelling barriers. Hence only for small contact areas and where the spin diffusion length is not very important for the device thin metal films can act as excellent injectors of spin into the device.
\begin{figure}[t]
\includegraphics[width=8.5cm, height=6.0cm]{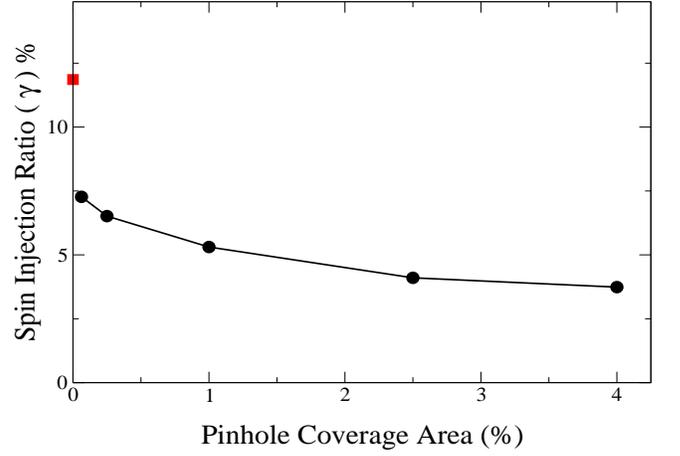}
\caption{(Color Online) Spin injection ratio as a function of the pinhole coverage area. Square (red online) shows the result for a perfect tunnelling barrier without pinholes. The total contact area for this device is $0.5 \mu$m $\times 0.5 \mu$m.}
\label{fig:pinholes}
\end{figure}

\indent Lastly we discuss pinholes in tunnelling barriers and its effect on the spin injection ratio. Although some discussions about pinholes\cite{buschow} have been made in the tunnelling magneto-resistance (TMR) experiments based on a simple resistor model by Oliver et~al. \cite{oliver}, there has been no work on this topic from the view point of the SDD model which is in general applicable to all spin-valve devices. In order to discuss pinholes we study a ferromagnet-semiconductor (FeNi-GaAs) interface with a single pinhole in the tunnelling barrier (AlO) as shown in Fig.~\ref{fig:devices} (c). The transport parameters used for GaAs are\cite{bala}: $\sigma_{\scriptscriptstyle{\rm{GaAs}}} = 10^{3} /\Omega$m, $\lambda_{\scriptscriptstyle{\rm{GaAs}}} = 1$ $\mu$m, $\alpha_{\scriptscriptstyle{\rm{GaAs}}} = 0$. Fig.~\ref{fig:pinholes} shows the effect of pinhole coverage area, i.e., the ratio of the pinhole area to the total contact area, on the spin injection ratio for the device. There are two competing effects here that determine the spin injection ratio, one of the conduction electrons passing through the pinhole (reducing $\gamma$) and second due to the tunnelling electrons passing via the tunnelling barrier (increasing $\gamma$). In order to understand the results obtained via simulations and thus the effect of pinholes, let us consider two channels one for the conduction electrons (pinhole channel) and the second for the tunnelling electrons (tunnelling barrier channel). By assuming that these channels are independent it can be shown that the spin injection ratio of the device comprising of these two channels can be given by,  
\begin{equation}
\gamma=\frac{\mathbf{j_{\scriptscriptstyle 1 ,\scriptscriptstyle \uparrow}}+\mathbf{j_{\scriptscriptstyle 2 ,\scriptscriptstyle \uparrow}}-\mathbf{j_{\scriptscriptstyle 1 ,\scriptscriptstyle \downarrow}}-\mathbf{j_{\scriptscriptstyle 2 ,\scriptscriptstyle \downarrow}}}{\mathbf{j_{\scriptscriptstyle 1 ,\scriptscriptstyle \uparrow}}+\mathbf{j_{\scriptscriptstyle 2 ,\scriptscriptstyle \uparrow}}+\mathbf{j_{\scriptscriptstyle 1 ,\scriptscriptstyle \downarrow}}+\mathbf{j_{\scriptscriptstyle 2 ,\scriptscriptstyle \downarrow}}}=\gamma_{\scriptscriptstyle 1}\frac{\mathbf{J_{\scriptscriptstyle 1}}}{\mathbf{J_{\scriptscriptstyle 1}}+\mathbf{J_{\scriptscriptstyle 2}}}+\gamma_{\scriptscriptstyle 2}\frac{\mathbf{J_{\scriptscriptstyle 2}}}{\mathbf{J_{\scriptscriptstyle 1}}+\mathbf{J_{\scriptscriptstyle 2}}} .
\label{eqn:15}
\end{equation}
where $\mathbf{j_{\scriptscriptstyle 1 ,\scriptscriptstyle 2 ,\scriptscriptstyle \uparrow ,\scriptscriptstyle \downarrow}}$, $\mathbf{J_{\scriptscriptstyle 1 ,\scriptscriptstyle 2}} = \mathbf{j_{\scriptscriptstyle 1 ,\scriptscriptstyle 2 ,\scriptscriptstyle \uparrow}} + \mathbf{j_{\scriptscriptstyle 1 ,\scriptscriptstyle 2 ,\scriptscriptstyle \downarrow}}$ are the up/down spin currents and total currents for the two channels and $\gamma_{\scriptscriptstyle 1, \scriptscriptstyle 2} = (\mathbf{j_{\scriptscriptstyle 1 ,\scriptscriptstyle 2 ,\scriptscriptstyle \uparrow}} - \mathbf{j_{\scriptscriptstyle 1 ,\scriptscriptstyle 2 ,\scriptscriptstyle \downarrow}})/(\mathbf{j_{\scriptscriptstyle 1 ,\scriptscriptstyle 2 ,\scriptscriptstyle \uparrow}} + \mathbf{j_{\scriptscriptstyle 1 ,\scriptscriptstyle 2 ,\scriptscriptstyle \downarrow}})$ are the spin injection ratios of the two channels.

Now for the case of the smallest pinhole if we consider the effect of area as shown previously (Fig.~\ref{fig:CA}) we obtain the $\gamma_{\scriptscriptstyle 1} = 7.38\%$. Using the result for perfect tunnelling barrier ($\gamma_{\scriptscriptstyle 2} = 11.85\%$) and the fact that the pinhole acts like a short circuit causing most of the current to pass through the pinhole ($\mathbf{J_{\scriptscriptstyle 1}}/(\mathbf{J_{\scriptscriptstyle 1}}+\mathbf{J_{\scriptscriptstyle 2}}) = 95\%$ from simulations) we get $\gamma = 7.6\%$ (from Eq.~(\ref{eqn:15})), which is quite close to the value $7.26\%$ obtained via simulations. The discrepancy of $0.34\%$ seen is due to the fact that the tunnelling barrier region now has a pinhole defect in it which was not considered while calculating $\gamma_{\scriptscriptstyle 2}$. This result cannot be understood using the 1D SDD model, which gives $\gamma = 0.6\%$, since the effect of area cannot be taken into account. Thus overall effect of pinholes is not just the sum of two individual 1D channels. We also simulate more than one pinhole to see if there is any correlation between the different pinholes. We observe no such correlations at distances of $\sim 300$ nm, $\sim 500$ nm and $\sim 800$ nm which are typically the experimentally observable distances for such spin valve devices. Hence we conclude that the spin injection ratio depends only on the effective coverage area of the pinholes and not the number of pinholes present in the tunnelling barrier.
\section{Conclusions}
In summary we present the results of a 3D SDD model and show that the 1D model fails to describe even the simple three terminal devices. We discuss the effects of N/SC height and contact area on the spin injection ratio of n-GaAs and show that direct contact spin injection is possible only for extremely small contact areas and height $\gg \lambda_{\scriptscriptstyle SC}$. We discuss the role of tunnelling barriers and show that thin metal films could be efficient spin injectors for small contact areas. Lastly the role of pinholes is discussed and we show that the spin injection ratio depends only on the effective area of the pinholes and no correlation between the number of pinholes and $\gamma$ is observed.
\section*{Acknowledgements}
We would like to thank Jos{\'e} Garc\'{\i}a-Palacios, Bijay Kumar Agarwalla and Meng Lee Leek for insightful discussions.


\begin{thebibliography}{*}
\bibitem{note0}
Since we are dealing with isotropic materials $\sigma$ should be treated here as a number.
\bibitem{note1}
Eq.~\ref{eqn:12} ensures that the spin current is conserved at the interface iff the interface is a constant potential surface. This assumption is valid for materials having the in plane conductivity component nearly equal to the out of plane component.
\bibitem{note3}
The thickness of the Cu layer as 50 nm is chosen purely due to computational limitations and in an experimental situation the thickness of the Cu layer should be made as thin as possible.
\end{thebibliography}
\end{document}